\documentclass[twocolumn,showpacs,prl]{revtex4}

\usepackage[utf8]{inputenc}
\usepackage[T1]{fontenc}
\usepackage[english]{babel} 
\usepackage{amsmath}
\usepackage{amsfonts}
\usepackage{amssymb}
\usepackage{amsthm}
\usepackage{euscript}
\usepackage{tikz}
\usepackage{url}
\usepackage{algorithm,algorithmic}

\newtheorem{theo}{Theorem}

\newtheorem{lemma}{Lemma}

\newcommand{\C}{\mathbb{C}}

\newcommand{\h}{\mathcal{H}}

\newcommand{\E}{{\EuScript E}}

\newcommand{\Prob}{{\mathbb P}}

\begin{document}

\title{Linear-Time Maximum Likelihood Decoding of Surface Codes over the Quantum Erasure Channel}

\makeatletter
\let\@fnsymbol\@arabic
\makeatother

\author{Nicolas Delfosse$^{1,2}$ and Gilles Zémor$^3$}

\affiliation{$^1$IQIM, California Institute of Technology, Pasadena, CA, USA\\
$^2$Department of Physics and Astronomy, University of California, Riverside, CA, USA\\
$^3$Mathematical Institute, IMB, UMR 5251, Bordeaux University, France}

\pacs{}

\begin{abstract}
Surface codes are among the best candidates to ensure the fault-tolerance
of a quantum computer. In order to avoid the accumulation of errors during a
computation, it is crucial to have at our disposal a fast decoding algorithm
to quickly identify and correct errors as soon as they occur.
We propose a linear-time maximum likelihood decoder for surface codes
over the quantum erasure channel. This decoding algorithm for dealing
with qubit loss
is optimal both in terms of performance and speed.
\end{abstract}

\maketitle

\vspace{.2cm}
\noindent
{\em Introduction---}
Surface codes \cite{kitaev2003:codes, dennis2002:memory} are one of 
the leading candidates to ensure the fault-tolerance of a quantum 
computer.
Error correction is based on the measurement of local operators on 
a lattice of qubits. The measurement outcome, called the syndrome, is 
then processed by the decoding algorithm which uses this information 
to infer the error which occurred.
In order to avoid the accumulation of errors during computation, 
it is essential for the decoder to be fast.
Any speed-up of the decoder leads indirectly to a reduction of the 
noise strength, since a shorter time between two rounds of correction
induces the appearance of fewer errors.

The quantum erasure channel \cite{grassl1997:erasure, bennett1997:erasure} 
is the noise model that represents photon loss or leakage outside the 
computational space in multi-level systems.
The loss of a qubit is equivalent to applying a random Pauli error to this qubit,
while giving, as additional data, the position of the error.
For stabilizer codes, this extra information reduces the decoding problem to 
solving a linear system, which can be done with cubic complexity.
In the particular case of surface codes, the syndrome of an error is a set of vertices
of a lattice and decoding amounts to finding a set of paths connecting these vertices by pairs.
One could pick two vertices and connect them by a path and repeat until
all the syndrome vertices are matched. This would lead to a quadratic complexity.
This strategy was adopted by Dennis {\em et al.} to decode Pauli errors \cite{dennis2002:memory}
or by Barrett and Stace in the case of a combination of Pauli errors and erasures \cite{barrett2010:loss}.

In the present work, we propose a linear-time maximum likelihood decoder
for erasures over surface codes.
This is optimal both in terms of performance and in terms of complexity.
Our algorithm can be used with any surface code, 
with arbitrary genus, and any type of boundary 
\cite{freedman2001:planar, bravyi1998:planar, delfosse2016:GSC}, 
including hyperbolic codes
\cite{freedman2002:systole, zemor2009:hyperbolic, breuckmann2016:hyperbolic}.
In comparison, in the case of Pauli errors the efficient algorithm for maximum likelihood 
decoding over surface codes obtained by Bravyi, Suchara and Vargo \cite{bravyi2014:MLD}, 
only applies to a restricted set of surfaces.


%

In the rest of the paper, we describe our decoding algorithm
and we prove that it is a maximum likelihood decoder.
To illustrate the decoding strategy, we first consider Kitaev's
surface codes, then we generalize the approach to surfaces 
with boundaries, which are more relevant for practical 
purposes \cite{freedman2001:planar, bravyi1998:planar, delfosse2016:GSC}.

\medskip
{\em Kitaev's surface codes --}
Kitaev's surface codes \cite{kitaev2003:codes} are obtained by imposing local constraints on qubits 
placed on a closed surface. 
Since only the combinatorial structure of the surface matters, we denote
by $(V, E, F)$ such a surface with vertex set $V$, edge set $E$ and
face set $F$. These three sets are assumed to be finite. An edge $e \in E$
is a pair of distinct vertices $e = \{u, v\}$. A face is a region of the surface
homeomorphic to a disc and delimited by a set of edges. 
We represent a face by the set of edges lying on its boundary.
We assume that the graph $(V, E)$ has neither loops nor multiple edges. 
We also suppose that its dual is well defined and satisfies the same properties.

Consider the Hilbert space $\h = (\C^2)^{E}$. Each qubit is indexed by an egde
$e$ and the Pauli operator acting on this qubit as the matrix $X, Y$ or $Z$
and acting trivially elsewhere is denoted respectively by 
$X_e, Y_e$ or $Z_e$.
{\em Kitaev's surface code} is defined to be the ground space of the Hamiltonian 
$$
- \sum_{v \in V} X_v - \sum_{f \in F} Z_f,
$$
where $X_v = \prod_{v \in e} X_e$ and $Z_f = \prod_{e \in f} Z_e$.
The operators $X_v$ and $Z_f$ generate a group $S$, the {\em stabilizer group},
which fixes the code space. Elements of $S$ are called {\em stabilizers}.
The $Z$-stabilizers, that are products of face operators $Z_f$,
are the operators of $\{I, Z\}^{\otimes E}$ whose support is a trivial cycle of $G$.
By {\em cycle}, we mean here a subset of edges of $G$ which 
meets every vertex an even number of times.
A cycle is said to be {\em trivial} if it lies on the boundary of a set of faces.
In the same way, $X$-stabilizers correspond to trivial cycles of the dual graph.
The correction of Pauli errors is based on the measurement of the generators $X_v$ and
$Z_f$ which tells us whether or not the error commutes with these operators.
The outcome of this measurement is called the {\em syndrome} of the error.
Errors with a trivial syndrome, meaning that commute with all the stabilizers, 
can be seen as operators acting on the code space and are called {\em logical operators}.
For instance, stabilizers are trivial logical operators. Non-trivial logical operators 
correspond to non-trivial cycles in the graph $G$ or its dual.

\medskip
{\em Maximum likelihood decoding for qubit loss --}
The {quantum erasure channel} is one of the most simple noise models. Each 
qubit is lost, or erased, independently with probability $p$. Such a loss can be detected and the
missing qubit is then replaced by a totally mixed state ${\mathcal I}/2$.
Writing ${\mathcal I}/2 = \frac{1}{4} (\rho + X \rho X + Y \rho Y + Z \rho Z)$,
we see that this new qubit can be interpreted as the original state which
suffers from a Pauli error $I, X, Y$ or $Z$ chosen uniformly at random.
The set of lost qubits is denoted by $\E$. 
The encoded state is subjected to a random uniform Pauli error 
$P$ whose support is included in $\E$. Denote this condition by $P \subset \E$.

Just like when dealing with Pauli noise, one can then measure the stabilizer
generators $X_v$ and $Z_f$ and try to recover the error $P$ from its syndrome.
The main difference with Pauli channels is the additional knowledge 
of the erasure pattern $\E$.
Since operators of $S$ act trivially on the code space, 
the goal of the decoder is to identify the coset $P \cdot S$ of the error, 
knowing the set $\E$ and the syndrome $\sigma$ of $P$.
The optimal strategy, called {\em maximum likelihood decoding}, is to maximize
the conditional probability $\Prob(P\cdot S | \E, \sigma)$.

To illustrate how the knowledge of the erasure $\E$ simplifies the decoding problem,
assume that we found an error $\tilde P \subset \E$ whose syndrome matches 
$\sigma$. 
Both errors $P$ and $\tilde P$ have the same syndrome, hence 
$\tilde P$ and $P$ differ in a logical operator $L \subset \E$, trivial or not.
Due to the fact that errors $Q \subset \E$ are uniformly distributed,
$\Prob(Q\cdot S | \E, \sigma)$ is proportional to the number 
$|Q\cdot S \cap \E|$ of Pauli errors of that coset that are included in $\E$. 
This number depends only on the number $|S \cap \E|$ of stabilizers 
having support inside $\E$, which shows that all the cosets are equiprobable. 
Therefore, maximum likelihood decoding consists simply of returning an error coset 
$\tilde P \cdot S$ such that $P \subset \E$ and the syndrome of $P$ is
equal to a given $\sigma$.
This proves that
\begin{lemma} \label{lemma:MLD}
Given an erasure $\E \subset E$ for a surface code and a measured syndrome $\sigma$,
any coset $\tilde P \cdot S$ of a Pauli error $\tilde P \subset \E$ of syndrome $\sigma$
is a most likely coset.
\end{lemma}
The same argument can be applied to any stabilizer code.

\medskip
{\em A linear-time maximum likelihood decoder --}
We now propose a fast algorithm that returns such a most likely 
coset for Kitaev's surface codes.
We detail the construction of the $Z$-part of the error.
The same algorithm will be applied to the dual graph to 
recover the $X$-part of the error.

Only measurements of operators $X_v$ can detect a $Z$-error.
The syndrome of a $Z$-error $P$ is thus the subset $\sigma(P) \subset V$ 
of vertices $v$ such that $X_v$ anti-commutes with this error.
Equivalently, it is the set of vertices surrounded by an odd number of
qubits supporting an error $Z$.
In order to translate our decoding problem into a graphical language,
denote by $\partial(A)$ the set of vertices that a subset $A \subset E$
encounters an odd number of times and call it the {\em boundary} 
of $A$. The syndrome of the $Z$-error pattern supported on $A$ is
exactly $\partial(A)$.
Given $\E \subset E$ and $\sigma \subset V$, we are looking for a subset 
of edges $A \subset \E$ such that $\partial(A) = \sigma$.

\begin{figure}[htbp]
\centering
\includegraphics[scale=.55]{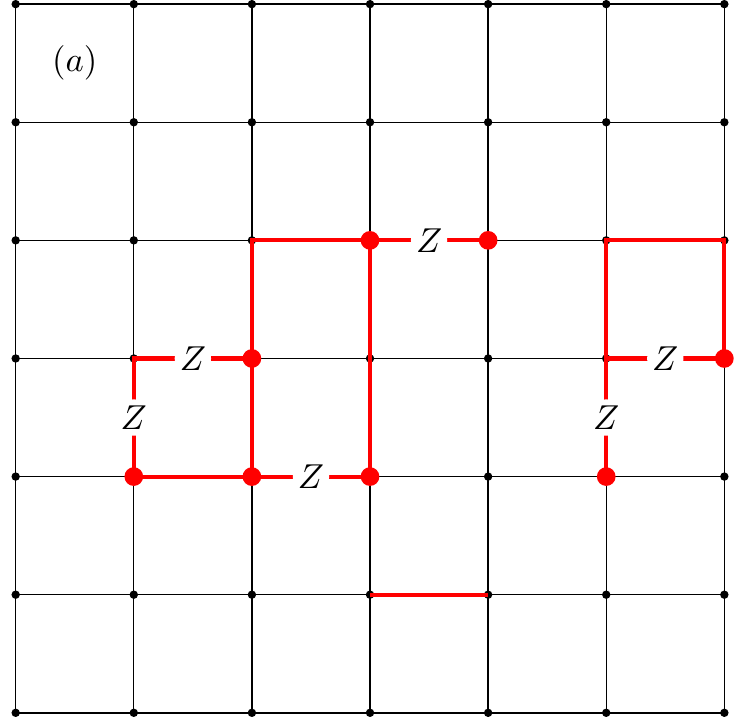}
\hspace{.05cm}
\includegraphics[scale=.55]{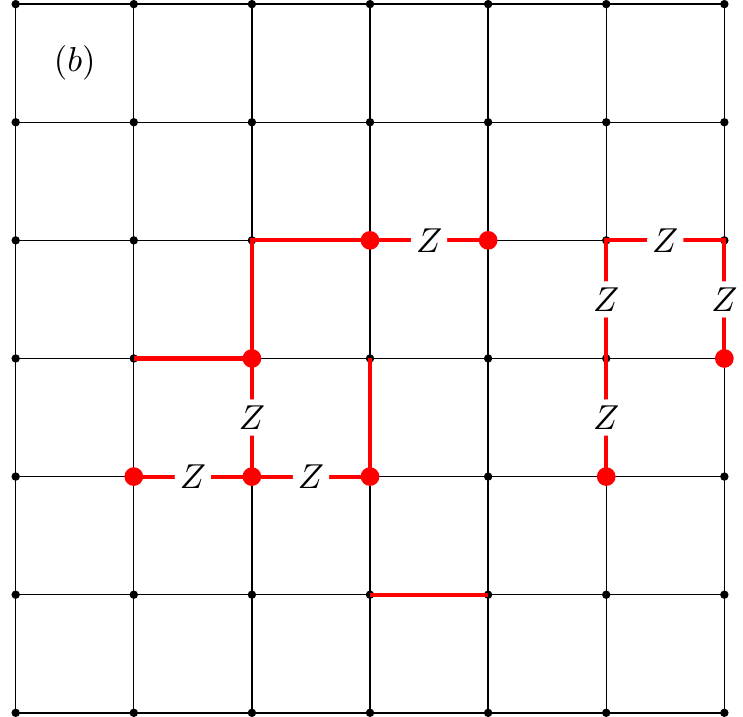}

	\caption{
	(a) A square lattice of the torus. Opposite sides are identified.
	Red thick edges mark the set $\E$ of erased qubits which support some
 	$Z$-error. Its syndrome is indicated by large red nodes.
	(b) A spanning forest $F_\E$ (thick red lines in (b)) is constructed.
	Then, starting from the leaves, an error included in the $F_\E$
	is constructed using the syndrome. Here, this provides a correct 
	estimation of the error up to a stabilizer.
	}
	\label{fig:decoding}
\end{figure}

We now describe Algorithm~\ref{algo:MLD} which is illustrated on
Fig.~\ref{fig:decoding}.

Paradoxically, an obstacle to a linear-time complexity is the presence of 
cycles in $\E$. Although cycles increase the number of paths from a
vertex to another and potentially make it easier to find one, they also
make it easier to make suboptimal choices.
Our basic idea is not to try to sequentially find paths that pair
the syndrome vertices together but instead to shrink recursively the
set of edges on which we have yet to make a decision.
To this end we select a {\em spanning forest} $F_\E$ inside $\E$, 
that is a maximal subset of edges of $\E$ that contains no cycle and spans all 
the vertices of $\E$. If $\E$ is a connected graph, then $F_\E$ is also 
connected and is called a {\em spanning tree}.
Such a forest can be found in linear time. 

Equipped with the forest $F_\E$ that contains all the syndrome vertices,
we can now find the required subset $A$ very efficiently. 
Starting with the empty set, we construct $A$,  
by applying recursively the following rules.

(R1) Pick a {\em leaf}, that is an edge $e = \{u, v\}$ connected to the forest through 
only one of its 2 endpoints, say $v$. The vertex $u$ is called a {\em pendant vertex}.
Assume first that $u \in \sigma$, then we add the edge $e$ to the set $A$ and we {\em flip} the 
vertex $v$. By flipping, we mean that $v$ is added to the set $\sigma$ if $v \notin \sigma$ 
and it is removed from $\sigma$ in the case $v\in \sigma$.
Then, $e$ is removed from the forest $F_\E$.

(R2) In the case when $u \notin \sigma$, this edge is simply removed from $F_\E$
and $A$ is kept unchanged.

Through these 2 steps, we peel the forest $F_\E$ until only an empty set
remains. The construction of the set $A$ is then complete.
This procedure relies on the following obvious remark, stated as a lemma to
emphasize the role of the two rules applied in Algorithm~\ref{algo:MLD}.
\begin{lemma}[\bf leaf alternative] \label{lemma:rules_alternative}
Let $A$ be a subset of edges of a tree $T$.
If $e = \{u, v\}$ is a leaf with pendant vertex $u$, then 
(R1) either $u \in \partial(A)$ and $e \in A$,
(R2) or $u \notin \partial(A)$ and $e \notin A$.
\end{lemma}

This strategy is guaranteed to end after a finite number of steps. 
It remains to show that it returns the expected set $A$.
We must verify that such a set $A$ exists 
and that the peeling process does not depend on the order in which leaves 
of the forest are removed.
This is done in the proof of Theorem~\ref{theo:proof_algo}.

\begin{algorithm}
\caption{Maximum Likelihood decoding}
\label{algo:MLD}

\begin{algorithmic}[1]
\REQUIRE A surface $G = (V, E, F)$, an erasure $\E \subset E$ and the syndrome $\sigma \subset V$
of a $Z$-error.
\ENSURE A $Z$-error $P$ such that $P \subset \E$ and $\sigma(P) = \sigma$.
\STATE Construct a spanning forest $F_\E$ of $\E$.
\STATE Initialize $A$ by $A = \emptyset$.
\STATE While $F_\E \neq \emptyset$, pick a leaf edge $e = \{u, v\}$ with 
pendant vertex $u$, remove $e$ from $F_\E$ and apply the 2 rules:
\STATE \hspace{.5cm}(R1) If $u \in \sigma$, add $e$ to $A$, remove $u$ from $\sigma$ 
and flip $v$ in $\sigma$.
\STATE \hspace{.5cm}(R2) If $u \notin \sigma$ do nothing.
\STATE Return $P = \prod_{e \in A} Z_e$.
\end{algorithmic}
\end{algorithm}

\begin{theo} \label{theo:proof_algo}
For surface codes with bounded degree and faces of bounded size,
applying Algorithm~\ref{algo:MLD} to the graph and to its dual produces 
a linear-time maximum likelihood decoder.
\end{theo}


During step 3 of the algorithm, a naive approach would be to 
look for a leaf by running over the forest at each round but this
strategy would lead to a super-linear complexity.
However, we can ensure linear complexity by running over the whole
forest and precomputing a list of leaves. 
For a bounded degree graph, this list can then be updated in constant time
at each round when an edge is removed from the forest.

\begin{proof}
Finding a spanning forest has a linear cost, then our algorithm 
runs over each edge of the forest only once, leading to a linear-time complexity
overall.
We have to prove that the set $A$, constructed by this algorithm,
satisfies the claimed properties. The fact that $A \subset \E$
is immediate. Only the condition $\partial(A) = \sigma$ deserves 
some attention.
First, we will show that, for any choice of $F_\E$, there exists a set
$A \subset F_\E$ such that $\partial(A) = \sigma$ and that this set is 
unique. 
Then we will see that applying (R1) and (R2), starting from the 
leaves, indeed constructs this set $A$.

There exists a subset $B$ such that $B \subset \E$
and $\partial(B) = \sigma$ since $\sigma$ is the syndrome of an
error. We will reroute the paths contained in $B$ to squeeze this 
subset inside $F_\E$ without changing its boundary.
Let $x_1, \dots, x_m$ be the edges of $B \backslash F_\E$.
By maximality of the forest $F_\E$, adding any extra edge $x_i$ to $F_\E$ 
creates a cycle $\gamma_{i} \subset F_\E \cup \{x_i\}$.
In order to remove $x_1$ from the set $B$, replace $B$ by 
$B_1 = B \Delta \gamma_{1}$ where $\Delta$ denotes
the symmetric difference of these two sets of edges. 
Then, $x_1 \notin B_1$, only edges of $F_\E$ are added to $B$
and $x_2, \dots, x_m$ are untouched.
By repeating this transformation, one creates a sequence 
$B_{i+1} = B_i \Delta \gamma_{i}$ such that 
$B_{i+1} \subset T_\E \cup \{x_i, \dots, x_m\}$ for $i=1, \dots, m$.
The last set, $B_m$, is included in $F_\E$.
Taking the symmetric difference with a cycle $\gamma_i$ preserves 
the boundary, {\em i.e.} $\partial(B_i) = \partial(B)$ for all $i$. 
This proves that the set $B_m$ satisfies both conditions $B_m \subset F_\E$ 
and $\partial(B_m) = \sigma$. This is our set $A$.

This set $A$ is unique.  
Indeed if there exists two such subsets $A$ and $A'$,
their symmetric difference $A \Delta A'$ is a subset of the forest 
which has a trivial boundary $\partial(A \Delta A') = \emptyset$
meaning that $A \Delta A'$ is a cycle.
Since this cycle is in a forest, it can only be the empty set, proving that $A = A'$.

Now that existence and unicity of $A$ are established, we see that 
the alternative offered by Lemma~\ref{lemma:rules_alternative} can
only end with the set $A$. 
The result of our algorithm is independent of the order in which we pick the 
leaves in step 3 by unicity of $A$.
The existence of $A$ garanties that our algorithm finds this set after peeling 
the whole forest.
\end{proof}

\medskip
{\em Surfaces with boundaries --}
Kitaev's construction of surface codes can be generalized to surfaces
with boundaries, that is closed surfaces punctured with holes. 
Introducing boundaries leads to a key simplification for the experimental 
realization of topological codes. One can obtain non-trivial surface codes
based on planar lattices.
This motivates the generalization of our decoding algorithm to such 
surface codes.
Two kinds of codes based on surfaces with boundaries have been suggested.
First, Freedman and Meyer noticed that one can consider a surface with 
boundaries \cite{freedman2001:planar}.
Algorithm~\ref{algo:MLD} can be immediately adapted to these codes.
Secondly, Bravyi and Kitaev introduced two different types of stabilizers 
supported on two types of boundaries \cite{bravyi1998:planar}.
Adapting our decoding algorithm to these codes presents 
two difficulties. First, the syndrome depends on the type of boundary and second, 
the spanning forest has to be grown in a way that depends on the boundary type.

We use the formalism of \cite{delfosse2016:GSC} 
that encompasses both generalizations of Kitaev's codes.
We consider a {\em surface $G = (V, E, F)$ with boundary}, which means that some
edges belong to a unique face. On the boundary, some edges and their endpoints 
are declared to be {\em open}. 
We denoted by $\partial_O E$ (resp. $\partial_O V$) these open sets
and by $\mathring V = V \backslash \partial_O V$ and  $\mathring E = E \backslash \partial_O E$
the non-open sets.
Qubits are placed on non-open edges and the {\em generalized surface code}
is defined as the ground space of the Hamiltonian 
$$
- \sum_{v \in \mathring V} X_v - \sum_{f \in F} Z_f
$$
where $X_v = \prod_{v \in e, e \in \mathring E} X_e$
and $Z_f = \prod_{e \in f, e \in \mathring E} Z_e$.
No qubit is placed on an open edge and open vertices do not support 
any operator $X_v$.

\begin{algorithm}
\caption{Maximum Likelihood decoding for surfaces with boundaries}
\label{algo:MLD_boundaries}

\begin{algorithmic}[1]
\REQUIRE A surface $G = (V, E, F)$ with open and closed boundaries, 
an erasure $\E \subset \mathring{E}$ and the syndrome $\sigma \subset \mathring{V}$
of a $Z$-error.
\ENSURE A $Z$-error $P$ such that $P \subset \E$ and $\sigma(P) = \sigma$.

\STATE Construct a spanning forest $F_\E$ of $\E$ with seed $\partial_O V \cap V(\E)$.
\STATE Initialize $A$ by $A = \emptyset$.
\STATE While $F_\E \neq \emptyset$, pick a leaf edge $e = \{u, v\}$ with 
pendant vertex $u \in \mathring{V}$, remove $e$ from $F_\E$ and apply the 2 rules:
\STATE \hspace{.5cm}(R1) If $u \in \sigma$, add $e$ to $A$, remove $u$ from $\sigma$ 
and flip $v$ in $\sigma$.
\STATE \hspace{.5cm}(R2) If $u \notin \sigma$ do nothing.
\STATE Return $P = \prod_{e \in A} Z_e$.
\end{algorithmic}
\end{algorithm}

\begin{figure}[htbp]
\centering
\includegraphics[scale=.55]{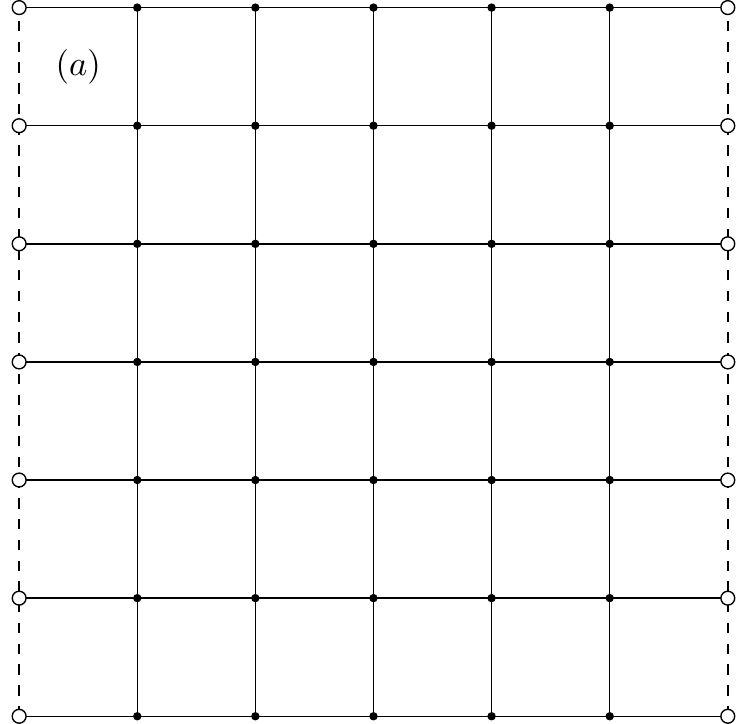}
\hspace{.05cm}
\includegraphics[scale=.55]{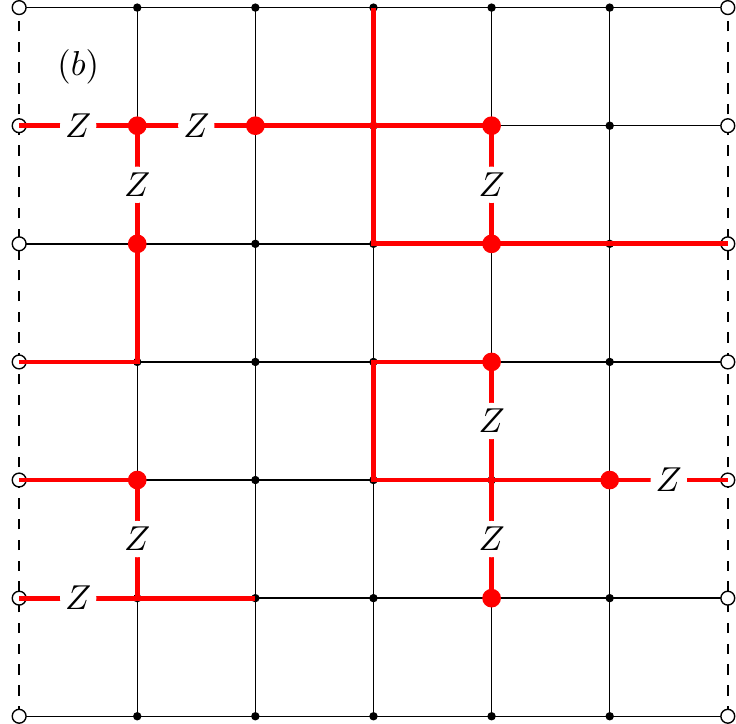}

\vspace{.4cm}
\includegraphics[scale=.55]{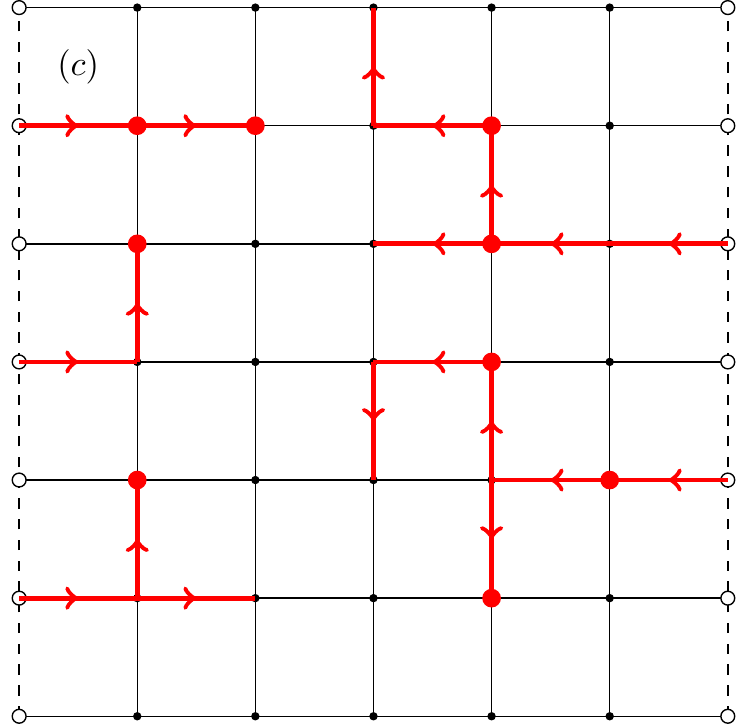}
\hspace{.05cm}
\includegraphics[scale=.55]{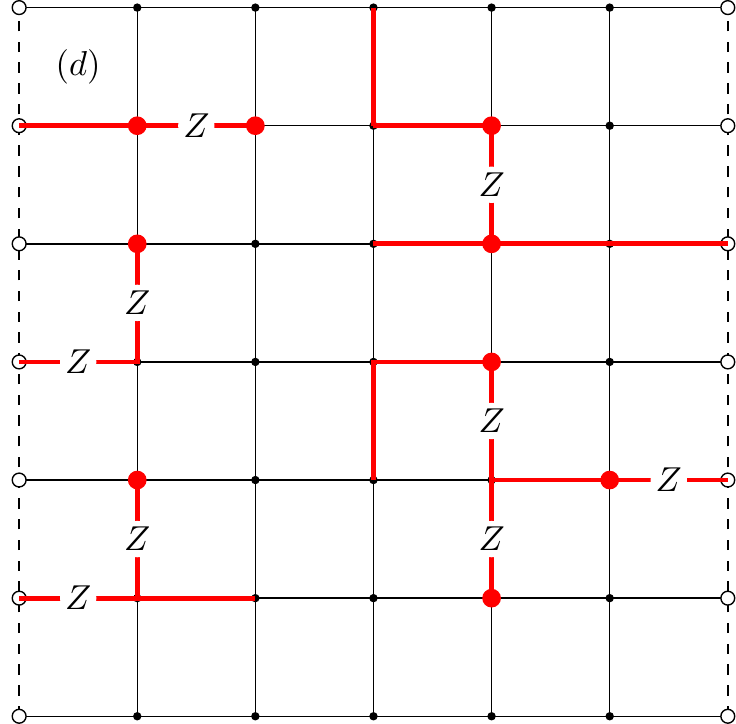}
	\caption{
		(a) Bravyi and Kitaev's code with open and closed boundaries.
	White nodes and dashed lines represent open vertices and open edges  .
		(b) Red thick lines indicate an erasure $\E$ with a $Z$-error and its syndrome which 
	is given by the large red vertices.	
		(c) A spanning forest $F_\E$, with open vertices as a seed. 
		Arrows show the way the forest is grown.
		(d) The error is estimated by reversing the arrows.	
	Our algorithm succeeds in identifying the error up to a stabilizer but 
	the choice of another forest may have produced a wrong estimation of the error.
	}
	\label{fig:decoding_boundaries}
\end{figure}

Consider an erasure $\E \subset \mathring E$
which comes with a Pauli error affecting erased qubits. 
Again, it suffices to focus on the correction of the $Z$-part of the error. 
Open vertices do not support any measurement $X_v$. Hence, the syndrome
of a $Z$-error of support $A \subset \mathring E$ is given by the restriction 
of $\partial(A)$ to non-open vertices. Denote by
$\mathring \partial(A) \subset \mathring V$ this restricted boundary.
The missing information on open vertices makes it impossible
to reconstruct the error starting from those vertices.
We must find a way to peel the whole forest using only non-open vertices.
In order to be sure that the peeling algorithm is not stuck before removing all
the edges of the forest, we will grow the forest starting from open vertices 
and peel it the other way round as depicted in Figure~\ref{fig:decoding_boundaries}.

Let us explain Algorithm~\ref{algo:MLD_boundaries}. An example is depicted
in Fig.~\ref{fig:decoding_boundaries}.
We must adapt the way the spanning forest is obtained.
First, let us explain a simple strategy to find a spanning tree of a connected graph $H = (V, E)$.
For a general graph, applying this method to all the connected components 
produces a spanning forest.
Our starting point is a tree $T$ that contains only a single arbitrary vertex $v$ of $H$
and no edge.
We grow $T$ by adding edges incident to the tree that connect $T$ with 
a vertex of $H$ that does not already belong to $T$. 
After adding $|V|-1$ edges, one gets our spanning tree.

In algorithm~\ref{algo:MLD_boundaries}, we will grow a spanning forest
of a graph $H = (V, E)$ equipped with a marked subset of vertices $O \subset V$
that we call the {\em seed}.
The spanning tree of a connected component containing 
a vertex $v_O \in O$ is constructed starting with this vertex $v_O$. Then, 
just as before we add edges that reach new vertices but we also require that 
these newly reached vertices do not belong to $O$.
If the connected component does not contain any seed vertex, 
the previous method applies.

\begin{theo} \label{theo:proof_algo_boundary}
For  generalized surface codes with bounded degree and faces of bounded size,
applying Algorithm~\ref{algo:MLD_boundaries} to the graph and to its dual produces 
a linear-time maximum likelihood decoder.
\end{theo}

\begin{proof}
Existence and uniqueness of the set $A$ follow from the same argument as in the proof of 
Theorem~\ref{theo:proof_algo} after replacing cycles by relative cycles. 
Recall that a relative cycle is a subset of edges that meets
each non-open vertex an even number of time. The space of relative cycles of graph 
is studied for instance in Section 4.1 of \cite{delfosse2016:GSC}.

Then, Lemma~\ref{lemma:rules_alternative}, which provides the recursive construction 
of the error, is used in an identical way.
We only need to make sure that the pendant vertices $u$ picked in step 3 are not open.
Our algorithm picks these vertices by reversing the construction of the forest with
open vertices as a seed. This guarantees that one can peel the whole forest and we 
end up with the correct set $A$ which provides the support of this error.
\end{proof}

{\em Concluding remarks --}
In this work, we considered the decoding problem of surface codes over 
the quantum erasure channel.
Despite the presence of inconvenient short cycles, we managed to design an 
optimal decoding algorithm that runs in linear time. 
Our basic idea is to remove these short cycles by decoding within a spanning 
forest. 
In the case of classical error correction, studying linear-time
decoding from erasures paved the way for better and better
linear (or quasi-linear) decoders in the case of more complicated channels.
We may hope that in the quantum setting, solving the decoding
problem for the erasure channel may similarly lead towards improved
decoders for more
complicated noise models and other families of codes.
In particular, a serious obstacle to decoding quantum LDPC codes is also
the presence of short cycles in their Tanner graph.
How to deal with them in general remains widely open
\cite{mackay2004:QLDPC, poulin2008:QLDPC, delfosse2013:LDPC, delfosse2014:correlations}.
It is crucial to consider such generalizations that may allow for fault-tolerant 
universal quantum computation with a considerably reduced overhead \cite{gottesman2014:LDPC}.
One could also consider the correction of losses assuming imperfect gates and measurements
\cite{whiteside2014:loss, suchara2015:leakage} or in the context of linear optical quantum computing
where photon losses are a major obstacle \cite{knill2001:LOQC, nielsen2004:LOCC_cluster, browne2005:LOCC_efficient, kieling2007:LOCC_percolation}.

\medskip
{\em Acknowledgement --}
ND thanks Jonas Anderson for his comments on a preliminary version of this work.
ND acknowledges funding provided by the Institute for Quantum Information and Matter, 
an NSF Physics Frontiers Center (NSF Grant PHY-1125565) with support of 
the Gordon and Betty Moore Foundation (GBMF-2644).


\begin{thebibliography}{23}
\expandafter\ifx\csname natexlab\endcsname\relax\def\natexlab#1{#1}\fi
\expandafter\ifx\csname bibnamefont\endcsname\relax
  \def\bibnamefont#1{#1}\fi
\expandafter\ifx\csname bibfnamefont\endcsname\relax
  \def\bibfnamefont#1{#1}\fi
\expandafter\ifx\csname citenamefont\endcsname\relax
  \def\citenamefont#1{#1}\fi
\expandafter\ifx\csname url\endcsname\relax
  \def\url#1{\texttt{#1}}\fi
\expandafter\ifx\csname urlprefix\endcsname\relax\def\urlprefix{URL }\fi
\providecommand{\bibinfo}[2]{#2}
\providecommand{\eprint}[2][]{\url{#2}}

\bibitem[{\citenamefont{Kitaev}(2003)}]{kitaev2003:codes}
\bibinfo{author}{\bibfnamefont{A.~Y.} \bibnamefont{Kitaev}},
  \bibinfo{journal}{Annals of Physics} \textbf{\bibinfo{volume}{303}},
  \bibinfo{pages}{27} (\bibinfo{year}{2003}).

\bibitem[{\citenamefont{Dennis et~al.}(2002)\citenamefont{Dennis, Kitaev,
  Landahl, and Preskill}}]{dennis2002:memory}
\bibinfo{author}{\bibfnamefont{E.}~\bibnamefont{Dennis}},
  \bibinfo{author}{\bibfnamefont{A.}~\bibnamefont{Kitaev}},
  \bibinfo{author}{\bibfnamefont{A.}~\bibnamefont{Landahl}}, \bibnamefont{and}
  \bibinfo{author}{\bibfnamefont{J.}~\bibnamefont{Preskill}},
  \bibinfo{journal}{Journal of Mathematical Physics}
  \textbf{\bibinfo{volume}{43}}, \bibinfo{pages}{4452} (\bibinfo{year}{2002}).

\bibitem[{\citenamefont{Grassl et~al.}(1997)\citenamefont{Grassl, Beth, and
  Pellizzari}}]{grassl1997:erasure}
\bibinfo{author}{\bibfnamefont{M.}~\bibnamefont{Grassl}},
  \bibinfo{author}{\bibfnamefont{T.}~\bibnamefont{Beth}}, \bibnamefont{and}
  \bibinfo{author}{\bibfnamefont{T.}~\bibnamefont{Pellizzari}},
  \bibinfo{journal}{Physical Review A} \textbf{\bibinfo{volume}{56}},
  \bibinfo{pages}{33} (\bibinfo{year}{1997}).

\bibitem[{\citenamefont{Bennett et~al.}(1997)\citenamefont{Bennett, DiVincenzo,
  and Smolin}}]{bennett1997:erasure}
\bibinfo{author}{\bibfnamefont{C.}~\bibnamefont{Bennett}},
  \bibinfo{author}{\bibfnamefont{D.}~\bibnamefont{DiVincenzo}},
  \bibnamefont{and} \bibinfo{author}{\bibfnamefont{J.}~\bibnamefont{Smolin}},
  \bibinfo{journal}{Physical Review Letters} \textbf{\bibinfo{volume}{78}},
  \bibinfo{pages}{3217} (\bibinfo{year}{1997}).

\bibitem[{\citenamefont{Barrett and Stace}(2010)}]{barrett2010:loss}
\bibinfo{author}{\bibfnamefont{S.~D.} \bibnamefont{Barrett}} \bibnamefont{and}
  \bibinfo{author}{\bibfnamefont{T.~M.} \bibnamefont{Stace}},
  \bibinfo{journal}{Physical review letters} \textbf{\bibinfo{volume}{105}},
  \bibinfo{pages}{200502} (\bibinfo{year}{2010}).

\bibitem[{\citenamefont{Freedman and Meyer}(2001)}]{freedman2001:planar}
\bibinfo{author}{\bibfnamefont{M.~H.} \bibnamefont{Freedman}} \bibnamefont{and}
  \bibinfo{author}{\bibfnamefont{D.~A.} \bibnamefont{Meyer}},
  \bibinfo{journal}{Foundations of Computational Mathematics}
  \textbf{\bibinfo{volume}{1}}, \bibinfo{pages}{325} (\bibinfo{year}{2001}).

\bibitem[{\citenamefont{Bravyi and Kitaev}(1998)}]{bravyi1998:planar}
\bibinfo{author}{\bibfnamefont{S.~B.} \bibnamefont{Bravyi}} \bibnamefont{and}
  \bibinfo{author}{\bibfnamefont{A.~Y.} \bibnamefont{Kitaev}}
  (\bibinfo{year}{1998}), \bibinfo{note}{arXiv:9811052}.

\bibitem[{\citenamefont{Delfosse et~al.}(2016)\citenamefont{Delfosse, Iyer, and
  Poulin}}]{delfosse2016:GSC}
\bibinfo{author}{\bibfnamefont{N.}~\bibnamefont{Delfosse}},
  \bibinfo{author}{\bibfnamefont{P.}~\bibnamefont{Iyer}}, \bibnamefont{and}
  \bibinfo{author}{\bibfnamefont{D.}~\bibnamefont{Poulin}},
  \bibinfo{journal}{arXiv preprint arXiv:1606.07116}  (\bibinfo{year}{2016}).

\bibitem[{\citenamefont{Freedman et~al.}(2002)\citenamefont{Freedman, Meyer,
  and Luo}}]{freedman2002:systole}
\bibinfo{author}{\bibfnamefont{M.~H.} \bibnamefont{Freedman}},
  \bibinfo{author}{\bibfnamefont{D.~A.} \bibnamefont{Meyer}}, \bibnamefont{and}
  \bibinfo{author}{\bibfnamefont{F.}~\bibnamefont{Luo}},
  \bibinfo{journal}{Mathematics of Quantum Computation, Chapman \& Hall/CRC}
  pp. \bibinfo{pages}{287--320} (\bibinfo{year}{2002}).

\bibitem[{\citenamefont{Z{\'e}mor}(2009)}]{zemor2009:hyperbolic}
\bibinfo{author}{\bibfnamefont{G.}~\bibnamefont{Z{\'e}mor}}, in
  \emph{\bibinfo{booktitle}{Proc. of the 2nd International Workshop on Coding
  and Cryptology, IWCC 2009}} (\bibinfo{organization}{Springer-Verlag},
  \bibinfo{year}{2009}), pp. \bibinfo{pages}{259--273}.

\bibitem[{\citenamefont{Breuckmann and
  Terhal}(2016)}]{breuckmann2016:hyperbolic}
\bibinfo{author}{\bibfnamefont{N.~P.} \bibnamefont{Breuckmann}}
  \bibnamefont{and} \bibinfo{author}{\bibfnamefont{B.~M.}
  \bibnamefont{Terhal}}, \bibinfo{journal}{IEEE Transactions on Information
  Theory} \textbf{\bibinfo{volume}{62}}, \bibinfo{pages}{3731}
  (\bibinfo{year}{2016}).

\bibitem[{\citenamefont{Bravyi et~al.}(2014)\citenamefont{Bravyi, Suchara, and
  Vargo}}]{bravyi2014:MLD}
\bibinfo{author}{\bibfnamefont{S.}~\bibnamefont{Bravyi}},
  \bibinfo{author}{\bibfnamefont{M.}~\bibnamefont{Suchara}}, \bibnamefont{and}
  \bibinfo{author}{\bibfnamefont{A.}~\bibnamefont{Vargo}},
  \bibinfo{journal}{Physical Review A} \textbf{\bibinfo{volume}{90}},
  \bibinfo{pages}{032326} (\bibinfo{year}{2014}).

\bibitem[{\citenamefont{MacKay et~al.}(2004)\citenamefont{MacKay, Mitchison,
  and McFadden}}]{mackay2004:QLDPC}
\bibinfo{author}{\bibfnamefont{D.~J.~C.} \bibnamefont{MacKay}},
  \bibinfo{author}{\bibfnamefont{G.}~\bibnamefont{Mitchison}},
  \bibnamefont{and} \bibinfo{author}{\bibfnamefont{P.~L.}
  \bibnamefont{McFadden}}, \bibinfo{journal}{IEEE Transaction on Information
  Theory} \textbf{\bibinfo{volume}{50}}, \bibinfo{pages}{2315}
  (\bibinfo{year}{2004}).

\bibitem[{\citenamefont{Poulin and Chung}(2008)}]{poulin2008:QLDPC}
\bibinfo{author}{\bibfnamefont{D.}~\bibnamefont{Poulin}} \bibnamefont{and}
  \bibinfo{author}{\bibfnamefont{Y.}~\bibnamefont{Chung}},
  \bibinfo{journal}{Quantum Information \& Computation}
  \textbf{\bibinfo{volume}{8}}, \bibinfo{pages}{987} (\bibinfo{year}{2008}).

\bibitem[{\citenamefont{Delfosse and Z{\'e}mor}(2013)}]{delfosse2013:LDPC}
\bibinfo{author}{\bibfnamefont{N.}~\bibnamefont{Delfosse}} \bibnamefont{and}
  \bibinfo{author}{\bibfnamefont{G.}~\bibnamefont{Z{\'e}mor}},
  \bibinfo{journal}{Quantum Information \& Computation}
  \textbf{\bibinfo{volume}{13}}, \bibinfo{pages}{793} (\bibinfo{year}{2013}).

\bibitem[{\citenamefont{Delfosse and
  Tillich}(2014)}]{delfosse2014:correlations}
\bibinfo{author}{\bibfnamefont{N.}~\bibnamefont{Delfosse}} \bibnamefont{and}
  \bibinfo{author}{\bibfnamefont{J.-P.} \bibnamefont{Tillich}}, in
  \emph{\bibinfo{booktitle}{2014 IEEE International Symposium on Information
  Theory}} (\bibinfo{organization}{IEEE}, \bibinfo{year}{2014}), pp.
  \bibinfo{pages}{1071--1075}.

\bibitem[{\citenamefont{Gottesman}(2014)}]{gottesman2014:LDPC}
\bibinfo{author}{\bibfnamefont{D.}~\bibnamefont{Gottesman}},
  \bibinfo{journal}{Quantum Information \& Computation}
  \textbf{\bibinfo{volume}{14}}, \bibinfo{pages}{1338} (\bibinfo{year}{2014}).

\bibitem[{\citenamefont{Whiteside and Fowler}(2014)}]{whiteside2014:loss}
\bibinfo{author}{\bibfnamefont{A.~C.} \bibnamefont{Whiteside}}
  \bibnamefont{and} \bibinfo{author}{\bibfnamefont{A.~G.}
  \bibnamefont{Fowler}}, \bibinfo{journal}{Phys. Rev. A}
  \textbf{\bibinfo{volume}{90}}, \bibinfo{pages}{052316}
  (\bibinfo{year}{2014}).

\bibitem[{\citenamefont{Suchara et~al.}(2015)\citenamefont{Suchara, Cross, and
  Gambetta}}]{suchara2015:leakage}
\bibinfo{author}{\bibfnamefont{M.}~\bibnamefont{Suchara}},
  \bibinfo{author}{\bibfnamefont{A.~W.} \bibnamefont{Cross}}, \bibnamefont{and}
  \bibinfo{author}{\bibfnamefont{J.~M.} \bibnamefont{Gambetta}}, in
  \emph{\bibinfo{booktitle}{Information Theory (ISIT), 2015 IEEE International
  Symposium on}} (\bibinfo{organization}{IEEE}, \bibinfo{year}{2015}), pp.
  \bibinfo{pages}{1119--1123}.

\bibitem[{\citenamefont{Knill et~al.}(2001)\citenamefont{Knill, Laflamme, and
  Milburn}}]{knill2001:LOQC}
\bibinfo{author}{\bibfnamefont{E.}~\bibnamefont{Knill}},
  \bibinfo{author}{\bibfnamefont{R.}~\bibnamefont{Laflamme}}, \bibnamefont{and}
  \bibinfo{author}{\bibfnamefont{G.~J.} \bibnamefont{Milburn}},
  \bibinfo{journal}{nature} \textbf{\bibinfo{volume}{409}}, \bibinfo{pages}{46}
  (\bibinfo{year}{2001}).

\bibitem[{\citenamefont{Nielsen}(2004)}]{nielsen2004:LOCC_cluster}
\bibinfo{author}{\bibfnamefont{M.~A.} \bibnamefont{Nielsen}},
  \bibinfo{journal}{Physical review letters} \textbf{\bibinfo{volume}{93}},
  \bibinfo{pages}{040503} (\bibinfo{year}{2004}).

\bibitem[{\citenamefont{Browne and Rudolph}(2005)}]{browne2005:LOCC_efficient}
\bibinfo{author}{\bibfnamefont{D.~E.} \bibnamefont{Browne}} \bibnamefont{and}
  \bibinfo{author}{\bibfnamefont{T.}~\bibnamefont{Rudolph}},
  \bibinfo{journal}{Physical Review Letters} \textbf{\bibinfo{volume}{95}},
  \bibinfo{pages}{010501} (\bibinfo{year}{2005}).

\bibitem[{\citenamefont{Kieling et~al.}(2007)\citenamefont{Kieling, Rudolph,
  and Eisert}}]{kieling2007:LOCC_percolation}
\bibinfo{author}{\bibfnamefont{K.}~\bibnamefont{Kieling}},
  \bibinfo{author}{\bibfnamefont{T.}~\bibnamefont{Rudolph}}, \bibnamefont{and}
  \bibinfo{author}{\bibfnamefont{J.}~\bibnamefont{Eisert}},
  \bibinfo{journal}{Physical Review Letters} \textbf{\bibinfo{volume}{99}},
  \bibinfo{pages}{130501} (\bibinfo{year}{2007}).

\end{thebibliography}

\newcommand{\SortNoop}[1]{}

\end{document}